\documentclass{article}

\usepackage{arxiv}
\usepackage[utf8]{inputenc} 
\usepackage[T1]{fontenc}    
\usepackage{hyperref}       
\usepackage{url}            
\usepackage{booktabs}       
\usepackage{amsfonts}       
\usepackage{nicefrac}       
\usepackage{microtype}      

\usepackage{graphicx}       

\title{Galaxy Learning - A Position Paper}

\author{
	Chao Wu\\
	Zhejiang University\\
	Hangzhou, China\\
  	\texttt{chao.wu@zju.edu.cn}\\
  	\And
  	Jun Xiao\\
  	Zhejiang University\\
  	Hangzhou, China\\
	\texttt{junx@cs.zju.edu.cn} \\
	\And
	Gang Huang\\
	Zhejiang Lab \\
	Hangzhou, China\\
	\texttt{huanggang@zju.edu.cn} \\
	\And
	Fei Wu\\
	Zhejiang University\\
	Hangzhou, China\\
	\texttt{wufei@zju.edu.cn} \\
}

\begin{document}
\maketitle

\begin{abstract}
The recent rapid development of artificial intelligence (AI, mainly driven by machine learning research, especially deep learning) has achieved phenomenal success in various applications. However, to further apply AI technologies in real-world context, several significant issues regarding the AI ecosystem should be addressed. We identify the main issues as data privacy, ownership, and exchange, which are difficult to be solved with the current centralized paradigm of machine learning training methodology. As a result, we propose a novel model training paradigm based on blockchain, named Galaxy Learning, which aims to train a model with distributed data and to reserve the data ownership for their owners. In this new paradigm, encrypted models are moved around instead, and are federated once trained. Model training, as well as the communication, is achieved with blockchain and its smart contracts. Pricing of training data is determined by its contribution, and therefore it is not about the exchange of data ownership. In this position paper, we describe the motivation, paradigm, design, and challenges as well as opportunities of Galaxy Learning.
\end{abstract}


\section{Motivation} \label{sec:motivation}
With an increasing amount of sensory data and advance of machine learning technologies, we are entering a new era, where our economy and business are shifting to a new data-driven approach: artificial intelligence (AI) is becoming the new engine for the rapid development of productivity, with (big) data as its fuel. With the expanding capability of deep neural networks, deep learning \cite{lecun2015deep,goodfellow2016deep} is becoming more data hungry, and therefore the importance of data has grown even more. Data have become a type of ``new money'' in the digital world \cite{shen2016pricing}.

However, in the current centralized machine learning paradigm where data are typically collected from end users or various sensors and uploaded to a remote server or a cluster of servers for data analysis and modelling, several significant issues exist, as follows:
\begin{itemize}
	\item \textbf{Privacy and security}: Current centralized machine learning methodologies have severe privacy issues, as users need to give out their valuable and/or sensitive data to third parties. In addition, it is risky to host data, especially users' data, on servers \cite{dillon2010cloud} as they can get hacked, and data can also be hacked when transferred from end users or sensors to servers. Current centralized machine learning paradigm suffers the risk of single point of failure as the entire system could be eventually being litigated out of existence or failing in spectacular crashes when the central server gets damaged abruptly. With the provision of the General Data Protection Regulation (GDPR) \cite{houser2018gdpr} and other data protection rules, data privacy and security merit more attention as a global socio-economic issue.
	\item \textbf{Ownership and Pricing}: Traditionally, data cannot be reasonably priced since its contribution to the model cannot be justified before being given out for modelling. However, user data include individual's privacy preferences and their personally identifiable information, and once users give out their data, they also give out the data ownership, which implies not only the possession of personal information but also the responsibility for personal information \cite{pantelis2013understanding}. As a result, although there exist various machine learning models, it is typically challenging to encourage individuals or organizations to contribute their data for modelling. With the value of data being more recognized, pricing mechanism is highly required in order to promote data collection.
	\item \textbf{Costs}: Modelling with a large amount of data incurs high costs, even on a cloud-based infrastructure \cite{aljarrah2015big}.   
	Model developers need to afford the high costs of the computation and storage for model training and its deployment. What is worse, datasets in realistic AI applications could range from TB-scale to PB-scale \cite{li2014communication}, making the current centralized machine learning methodologies not effective to complete modelling tasks in time. Also, data collection needs its own cost, and it is becoming higher with the value of data being more recognized and the regulation of data being increasingly stricter. Although big companies might afford the above costs, such high costs bring the barrier for individual developers and start-ups. 
\end{itemize}

Regarding the above challenges, Google proposed the concept of federated learning which trains a model while training data remain distributed \cite{konevcny2016federated,mcmahan2016communication,mcmahan2017federated}. Most recently, \cite{yang2019federated} extended the original federated learning to horizontal federated learning, vertical federated learning, and federated transfer learning. Although federated learning could boost the privacy of data and enhance the security of training, it is a partial solution because the users' privacy could still be undermined and the value of data could not be fairly priced. To further tackle the privacy protection issue, cryptography should be taken advantage of, and methods including multi-party computation (MPC) \cite{du2001secure}, differential privacy (DP) \cite{dwork2006differential}, and homomorphic encryption (HE) \cite{gentry2009fully} are available. As an attempt to price data for machine learning, \cite{kim2018on} leverages the idea of blockchain when performing federated learning. Unfortunately, a comprehensive solution to satisfyingly dealing with the above main issues all together is not yet available.

In order to build a new paradigm to encourage the exchange of data usage rather than data ownership and fully utilize the value of data in modelling, we propose to build a Galaxy Learning platform. In this distributed machine learning platform, instead of collecting data in centralized context, we leave data where they belong to and we move modelling towards them. Therefore, users' privacy and data ownership are preserved.  As the modelling is distributed, the paradigm will be robust against intervention by antagonists, whether legitimate governments or criminal elements. Moreover, single point of failure could be avoided as the paradigm is free of any central authority or point of control that can be attacked or corrupted. Cost of building a centralized computation infrastructure is distributed among participating nodes. In addition, the Galaxy Learning platform provides a pricing mechanism through which the contribution of data could be reasonably priced without violating the privacy and security of data. We believe this is a more realistic approach to tackle the issues we are confronted with when applying current centralized machine learning paradigm.

The Galaxy Learning platform has the following design considerations:
\begin{itemize}
	\item \textbf{Blockchain with smart contracts as decentralized modelling infrastructure}: Instead of deploying a big model on a server and gather the data, we deploy an initial model to a decentralized blockchain network with the form of smart contracts \cite{wood2014ethereum} and use the blockchain network and its computation facility (e.g., Ethereum Virtual Machine, EVM) as modelling infrastructure. The initial model is trained individually and aggregated by demand, for example, via federated learning, and the consensus mechanism is used to ensure the synchronization of modelling activities. The modelling infrastructure can be run on a wide range of computing devices, and we use blockchain to secure the ownership of digital assets. We train the model off-chain, and update it on-chain. 
	\item \textbf{Moving models to data with further protection}: By deploying models to blockchain, we move models towards data instead of moving data to models as in traditional approaches. Models are sent to data owners and trained with secure local access to encrypted data. As a result, the data owners can control their data and preserve their ownership and privacy. On the other hand, it is also necessary to protect the ownership of models, which mainly consist of the trained parameters. Therefore, we shall further protect models by applying advanced cryptographic techniques such as HE and MPC. 
	\item \textbf{Pricing data with its contribution to modelling}: A single modelling task is distributed into a number of smaller modelling tasks with different data providers and then the results (i.e., model updates from participatory data providers) are federated as a new (updated) model. The contribution of each data provider can be assessed by its contribution to the model update (i.e., the improvement to model accuracy), which can be used for pricing data. With this approach, the pricing of data is based on its usage, but not the exchange of its ownership.
\end{itemize}

In this position paper, targeting at designing a distributed machine learning platform with these considerations, we propose a new paradigm as well as the principle of system design, and discuss the key issues related. 

\section{Paradigm} \label{sec:paradigm}
In this section, we present a new paradigm, named Galaxy Learning, for machine learning training. The ideas discussed in Section \ref{sec:motivation} will all be covered in this distributed machine learning platform.

\subsection{Actors}
First of all, there are three roles within the Galaxy Learning platform:
\begin{enumerate}
	\item \textbf{Data provider}: Data providers (i.e., data owners, as shown in Figure \ref{fig:data_provider}) are typically end users (i.e., individuals, companies, or any organizations having their own data), willing to utilize their data for service exchange or other incentives. They collect data from sensors or other data sources. Before the data are contributed to modelling, data providers need to evaluate the quality of data and provide data validation as well as its schema. We need to emphasize that when we say ``data providers provide their data'', we mean the data are granted for being used rather than being given out to some other parties.
	\begin{figure}[!ht]
		\centering
		\includegraphics[scale=0.3]{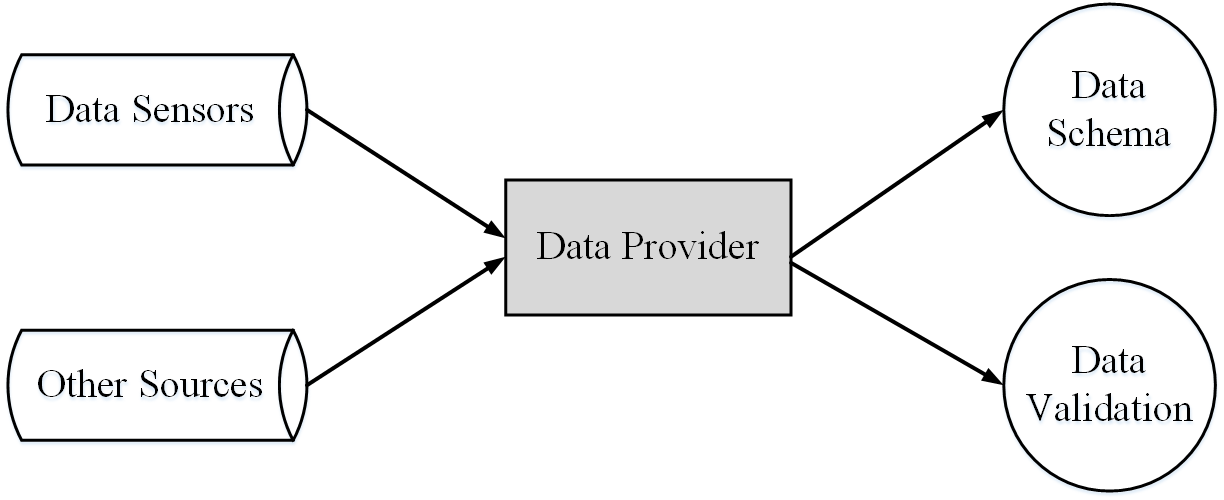}
		\caption{Data provider.\label{fig:data_provider}}
	\end{figure}
	\item \textbf{Model provider}: Model providers (i.e., data users or data requesters, as shown in Figure \ref{fig:model_provider}) utilize the data from data providers to develop machine learning models. A machine learning model is a function trained by a learning algorithm with training data. It can be an initial model without any training, or an existing model pre-trained by model providers. Such a model is typically developed by scientific researchers, industrial companies, or other model developers (e.g., open-source model developers). Model providers also act as a training task provider who initializes a model training task with data schema for required training data, test data to evaluate model update, and reward plan to pay data providers. In traditional paradigms, model providers need to set up training and deployment infrastructures for their models and collect their training data; in our distributed machine learning paradigm, they only need to provide their models as smart contracts to computation providers.
	\begin{figure}[!ht]
		\centering
		\includegraphics[scale=0.3]{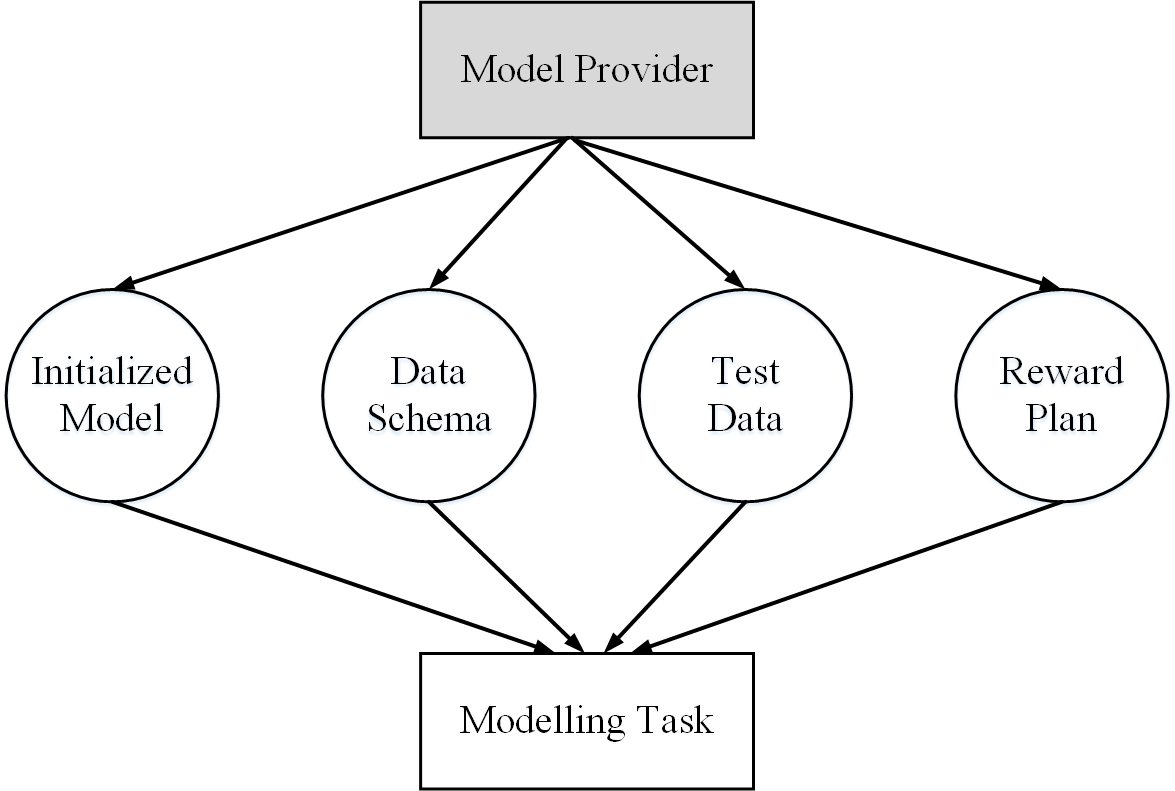}
		\caption{Model provider.\label{fig:model_provider}}
	\end{figure}
	\item \textbf{Computation provider}: Computation providers (as shown in Figure \ref{fig:computation_provider}) are nodes in the blockchain network to run the smart contracts for model training. They provide a secure and controlled training environment where both data and models are protected. They also measure the modelling contribution from the participating data. Generally, a model training task is distributed among multiple computation providers as a federated learning task. Please note that any node in the blockchain network can become a computation provider if it is willing to provide computation (ranging from high performance GPU clusters to mobile devices), even when it is also a data provider or model provider.
	\begin{figure}[!ht]
		\centering
		\includegraphics[scale=0.3]{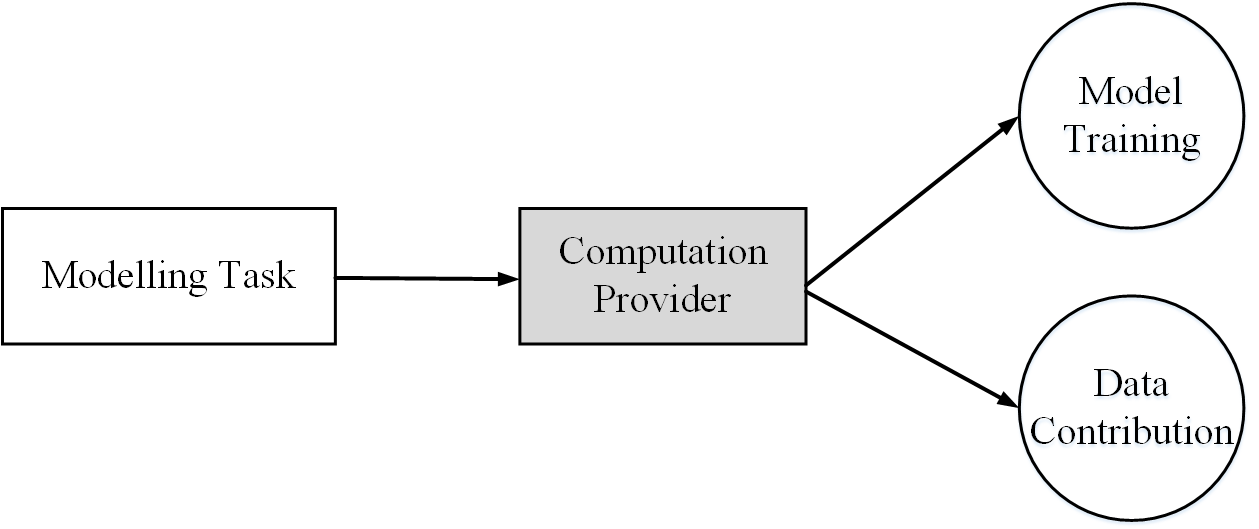}
		\caption{Computation provider.\label{fig:computation_provider}}
	\end{figure}
\end{enumerate}
In addition to the above actors, there are other nodes supporting the Galaxy Learning platform in the blockchain network: 1) \textbf{Notification node}, which takes the responsibility of communication and event triggers; 2) \textbf{Blockchain node}, which records the addresses of participating nodes and their models/data.  

\subsection{Components} \label{subsec:components}
We now present the main components in the Galaxy Learning paradigm:
\begin{enumerate}
    \item \textbf{Data storage}: Although data providers can choose any storage schema as they prefer, it needs a generic interface to access these data for modelling. We propose to use a peer-to-peer distributed file system (DFS, e.g., InterPlanetary File System, IPFS) with data schema as such an interface. User’s data after encryption is stored in their devices with the DFS as a versioned file system which can track versions of data over time. A data quality validation function is executed at user' local devices on the data to produce two outputs: 1) Quality score, which is an indicator for data quality according to certain measures (e.g., amount, variance); 2) Data signature, which is a hash key created for data to ensure the data used for quality validation is the same for the modelling. Encrypted data on DFS is coupled with a data schema, which describes meta information about the data (e.g., type, attributes). The data schema and quality score are used as matching criteria for modelling.
    \item \textbf{Model provisioning}: Model providers develop an initial machine learning model according to its requirement, in programming languages for smart contracts (e.g., Chaincode for Hyperledger, Solidity for Ethereum). In addition to the model, model providers also need to provide the following items to create a modelling task: 1) Data schema, which is used for training data matching; 2) Test data, which is used for performance evaluation; 3) Rewarding strategy, which is a plan to determine the contribution of data providers and their rewards. A modelling task will be published to the model repository on blockchain. The notification node will then notify potential data providers to initialize the model training.
    \item \textbf{Model training}: When a modelling task is initiated and published, a smart contract is executed to match the modelling task with data providers, according to the modelling requirement and data schema and its quality score. When matched with $N$ data providers, the task is distributed into $N$ sub-tasks, with an identical initial model for different training data. Federated learning is adopted here for distributed training, while the training task is taken on computation providers (e.g., EVM nodes in the Ethereum network). Encrypted data are transferred to computation providers. Also, we consider to protect model parameters during training with HE in order to protect the privacy of the model.
    \item \textbf{Pricing engine}: After training, the final model will be tested by test data, and if it achieves the performance criteria, reward plan is executed to reward data providers and computation providers. The reward for computation providers depends on the computation of smart contracts, which is similar to the Gas for Ethereum. The reward for data providers is determined by their contribution to the model update, with the strategy defined in the reward plan (i.e., the reward is divided according to the rank of accuracy improvement). The pricing engine divides the reward from model providers according to reward plan, and we note that an optimized plan can be estimated with game theory.
    \item \textbf{Model deployment}: We need to mention the main focus of Galaxy Learning is about model training, not its inference. A trained model can be deployed into or out of the blockchain network. If it is deployed into the blockchain network, corresponding smart contracts are required to define the interaction between the model and its consumers. A pricing smart contract can also be employed to define the price of model usage as well as the reward for computation providers and data providers. To establish a universal model inference environment, we had better utilize all available hardware to run the model. To achieve this, models, especially deep learning models, need to be compressed in order to reduce the size and accelerate its execution; on the other hand, we need to develop a universal modelling library so that a model can be run on any platform. 
\end{enumerate}

\section{Design} \label{sec:design}
\begin{figure}[!ht]
\centering
\includegraphics[scale=0.75]{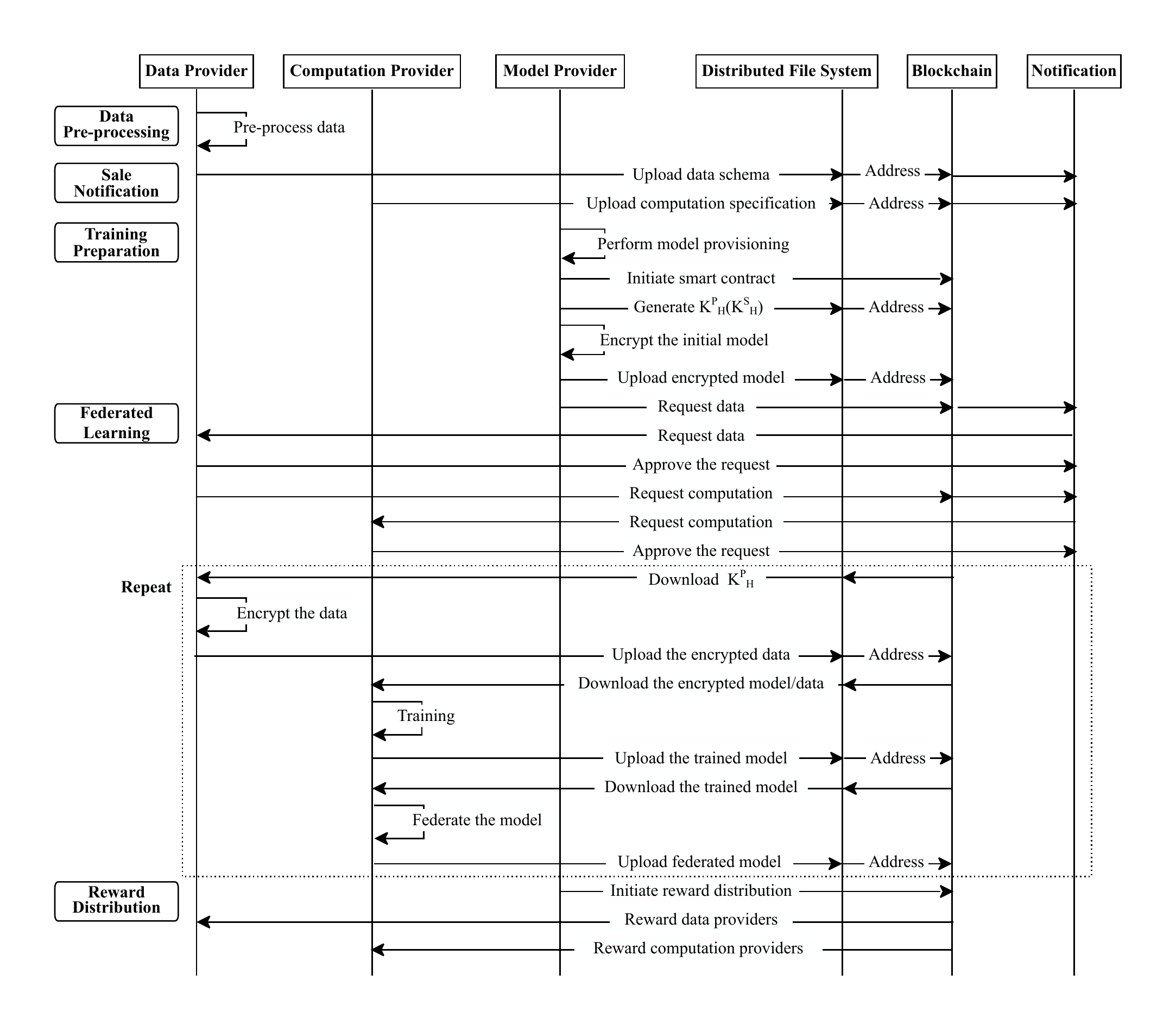} 
\caption{An illustration of the designed workflow. Herein, $K^P_H$ ($K^S_H$) represent the pair of homomorphic encryption keys.}\label{fig:workflow}
\end{figure}

In this section we describe a reference design of the Galaxy Learning platform on Ethereum \cite{wood2014ethereum}, which is a popular blockchain network that enables developers to build and deploy distributed applications. The whole workflow of the system is illustrated in Figure \ref{fig:workflow}, and all tokens used within the workflow are transferred according to smart contracts. Whisper\footnote{Ethereum Whisper: https://github.com/ethereum/wiki/wiki/Whisper}, which provides dark (plausible denial over perfect network traffic analysis) communications to two correspondents that know nothing of each other but a hash, is used to signal to each other in order to ultimately collaborate between nodes. 

\subsection{Data pre-processing}
Data pre-processing is performed offline locally by data providers. There are several sub-steps involved in this step:
\begin{enumerate}
    \item Sharding is performed to divide data into small portions, as it is difficult to scale with big chunks of data and some data consumers might only wish to purchase a small subset of data instead of everything. Sharding enables micropayments in the system.
    \item Quality scores are calculated using a commonly agreed quality measurement function. The quality measurement function can take inputs from multiple parties. When the function is simple, methods such as HE and MPC can be used to preserve privacy.
    \item Metadata (such as location, device id, data description, etc.) are prepared.
\end{enumerate}
    
\subsection{Sale notification}
Data providers and computation providers notify the blockchain network that they are willing to sell their data and computational ability. Specifically:
\begin{enumerate}
    \item Data providers firstly store data schema (including quality scores, metadata, and cost) on the distributed file system. The schema addresses are broadcasted to all nodes in the blockchain network. Any node then can locate and view schema using the corresponding address.
    \item Similarly, computation providers store the specification of their computing resources (including type of resource and cost) on the distributed file system. The specification addresses are also broadcasted to all nodes in the blockchain network. Any node in the network can locate and view the specification using the corresponding address.
\end{enumerate}
    
\subsection{Training preparation}
Model providers prepare the machine learning training by performing the following steps:
\begin{enumerate}
    \item As discussed in Section \ref{subsec:components}, model providers perform model provisioning and deploy the reward plan as a smart contract. The smart contract is then initiated on the blockchain.
    \item Model providers locally generate a pair of HE keys ($K^P_H$/$K^S_H$), which is partially opened to the blockchain network (i.e., the public key $K^P_H$ is written into smart contracts and broadcasted to all nodes). Note that we will never reveal the private key $K^S_H$.
    \item Model providers encrypt the initial model parameters $M_0$ using HE public key $K^P_H$. They then store the encrypted model $f_H(M_0)$ on the distributed file system, and uploads its hash to smart contracts.
\end{enumerate}
    
\subsection{Federated learning}
The following sub-steps are executed in sequence, and the training process will be repeated until the global model is converged:
\begin{enumerate}
    \item The model provider finds a suitable data resource for training and then asks for approval from the corresponding data provider. Note that the cost of data is specified in the corresponding schema.
    \item If data providers accept the request, they will search for the most suitable computation provider on the blockchain network and then ask for approval from it. Note that a rational data provider will only select computation provider that costs less than the token received from the model provider.
    \item If computation providers also accept the request, the training process starts:
    \begin{enumerate}
        \item Data providers download the HE public key $K^P_H$. Training data ($D_i$) are encrypted using HE public key. Encrypted data $f_H(D_i)$ will then be sent to the distributed file system.
        \item Computation providers download both the encrypted model $f_H(M_0)$ and encrypted data $f_H(D_i)$. At this point, computation providers perform training and the results ($f_H(M_i)$) will be saved in the distributed file system and uploaded to the network.
        \item Computation providers download certain result ($f_H(M_i)$). Then the results are aggregated into a global model for next round of federated learning.
    \end{enumerate}    
\end{enumerate}

\subsection{Reward distribution}
When the above federated learning process ends, the model provider calculates the contribution of each data provider and distributes the tokens in reward pool to these providers. The tokens in reward pool will also be distributed to computation providers, and this part of reward depends on the computation of smart contracts, which is similar to the Gas for Ethereum. In addition, all the HE-encrypted results will be visible to all participating parties, and the correctness of reward distribution can be verified off the chain.

\section{Challenges and Opportunities} \label{sec:issues}
We have provided the paradigm and design of the proposed Galaxy Learning platform. The fast development of machine learning, blockchain and cryptography opens up great opportunities for us, whereas the high complexity of the proposed platform poses great challenges. To successfully establish such a new AI ecosystem, we discuss several challenges and opportunities regarding the Galaxy Learning platform in this section.
\begin{itemize}
    \item \textbf{Performance}: To build a practical Galaxy Learning platform, we need to make sure its performance meets modelling requirement, especially considering the heavy modelling tasks associated with HE and MPC. The blockchain scalability (i.e., transactions per second) will become a key issue \cite{croman2016on,zheng2018blockchain} when the network is used in real-world applications. Although private blockchain and consortium blockchain could alleviate the problem with various new consensus mechanisms proposed \cite{baird2016hashgraph}, we need to find a solution to support trade-off between scalability and the level of decentralization. We also need to consider new security requirements in the distributed machine learning context \cite{papernot2018marauder}.
    \item \textbf{Interoperability}: Interoperability is important within and outside of the Galaxy Learning platform. Within Galaxy Learning, we need to support different blockchains and different devices (powerful/affordable, with/without GPU, etc.). For outside world, we need to make sure the Galaxy Learning interacts with traditional AI contexts. We adopted the principle of federated learning in the Galaxy Learning, but further research is required to extend it to more general situations rather than CNN \cite{mcmahan2016communication}. Moreover, it will be interesting to build a model update sharing mechanism or other serverless mechanism instead of model integration which is used in traditional federated learning.
    \item \textbf{Incentive}: The Galaxy Learning platform rewards data providers according to their contribution to the modelling. This is a new approach to price the data \cite{liang2018survey}, however, how to decide the incentive is not so obvious. Contributions from different data are dependent, and the contribution of data in modelling could go beyond the improvement on model accuracy. We are working on game theory to conduct the pricing, and we choose the improvement of model generalization as the utility function. Data providers could test each other's model update with their own data, and an incentive mechanism is required to help them decide which model update to adopt. The final model is determined by general consensus through data providers' actions.
\end{itemize}

At last, we need to mention that the proposed new paradigm of distributed machine learning is capable to couple with traditional centralized paradigm: a model can be initially trained in a centralized manner and then further trained in the distributed context. For example, a company can firstly train a basic classifier with its own data; while in traditional paradigm it will be difficult to further collect and feed external data to the classifier, now the company can distribute the (encrypted) model to the Galaxy Learning platform and gain a refined model with the data previously unavailable.

\bibliographystyle{unsrt}
\bibliography{main}

\end{document}